\documentclass[12pt]{iopart}
\usepackage{graphicx}
\usepackage{amsfonts}
\usepackage{amssymb}
\usepackage{mathptmx}

\begin{document}

\title{AC Josephson properties of phase slip lines in wide tin films}

\author{V.M.Dmitriev$\dag \ddag$, I.V.Zolochevskii$\dag$,
E.V.Bezuglyi$\dag$, D.S.Kondrashev$\dag$}

\address{
$\dag$ B.Verkin Institute for Low Temperature Physics and
Engineering, National Academy of Sciences of Ukraine, 61103
Kharkiv, Ukraine.}
\address{
$\ddag$ International Laboratory of High Magnetic Fields and Low
Temperatures, 95 Gajowicka St., 53-421 Wroclaw, Poland. }

\ead{dmitriev@ilt.kharkov.ua}

\begin{abstract}
Current steps in the current-voltage characteristics of wide
superconducting Sn films exposed to a microwave irradiation were
observed in the resistive state with phase slip lines. The
behaviour of the magnitude of the steps on the applied irradiation
power was found to be similar to that for the current steps in
narrow superconducting channels with phase slip centers and, to
some extent, for the Shapiro steps in Josephson junctions. This
provides evidence for the Josephson properties of the phase slip
lines in wide superconducting films and supports the assumption
about similarity between the processes of phase slip in wide and
narrow films.

\end{abstract}

\submitto{\SUST}


\maketitle

\section{Introduction}

The concept of phase slip lines (PSLs) formation by the transport
current in wide supercon\-ducting films as a mechanism of their
transition from the resistive vortex state to the normal state
\cite{Vol} suggests a similarity of the phase slip processes in
the PSLs and in phase slip centers (PSCs) in narrow
superconducting channels. It has been recognized (for a review,
see, e.g., \cite{Dmitrenko1,Dmitriev1}) that the initiation of
PSLs is responsible for occurrence of alternate voltage steps and
linear portions in the current-voltage characteristic (IVC) of
wide films, similar to the IVC features associated with PSCs in
narrow channels. Experimental investigations of the voltage
distribution in the vicinity of the PSL \cite{Dmitrenko} disclosed
that the variations of the quasiparticle potential spread over the
distance of the longitudinal electric field penetration depth
$l_{\rm E}$, which is also typical for a PSC. Theoretical results
of analytical \cite{Lempitskij} and numerical \cite{Weber}
investigations of the PSL structure largely support suggested
analogies between the PSLs and PSCs.

Important property of a PSC is oscillation of the order parameter
in the PSC core with the Josephson frequency (ac Josephson effect)
\cite{Skocpol,Tidecks}. The expected Josephson properties of PSLs
have been first examined in \cite{Sivakov}; however, in this
experiment, the vortex portion in the IVC of the sample was absent
which gives rise to a question about the interpretation of this
sample as a wide film. Indeed, the width of the sample used in
\cite{Sivakov} was 5~$\mu$m, whereas the penetration depth of the
normal-to-film magnetic field $\lambda_\perp$ was 3-5~$\mu$m,
i.e., close to the film width. As shown in \cite{Dmitriev1}, if
the film width $w$ does not exceed the quadrupled penetration
depth, i.e., at $w<4\lambda_\perp(T)$, the transition of the
superconducting film to the resistive phase slip state is similar
to that in a narrow channel: it occurs as soon as the transport
current approaches the value of the Ginzburg-Landau depairing
current, bypassing the stage of the moving vortex lattice. Hence,
strictly speaking, the experimental data in \cite{Sivakov} were
obtained for a narrow superconducting channel rather than for a
really wide film.

In the present paper, we report the results of a series of
experiments which demonstrate ac Josephson properties of PSL for
deliberately wide films with well pronounced initial vortex
portions in the IVC. Following the results of theoretical
\cite{Asla} and experimental \cite{Dmitriev1} investigations of
the stability of the vortex state in wide films, we assume the
absence of moving vortices in the phase slip state of the samples.

\section{Experimental results}

We investigate Sn films fabricated by a novel technique
\cite{Dmitriev1} which ensures minimum defects both at the film
edges and in its bulk. The IVCs of the samples were obtained by a
standard four-probe method. While measuring, the samples were
placed in a double screen of annealed permalloy. The parameters of
some films are listed in table 1. The electron mean free path due
to the scattering on impurities, $l_{\rm i}$, was evaluated by the
formula $l_{\rm i} = l_{\rm ph}(R_{300}/R_{4.2}-1)$
\cite{Imp1,Imp2}. Here $l_{\rm ph} = 9.5$ nm is electron-phonon
scattering length in Sn at room temperature \cite{Imp2},
$R_{300}$ is the film resistance at room
temperature and $R_{4.2}$ is the residual film resistance.
\begin{table}
\caption{Parameters of the film samples: $L$ is the length, $w$
the width, $d$ the thickness of the sample, and $l_{\rm i}$ is the
electron mean free path.} \label{tab}
\begin{center}
\begin{tabular}{|c|c|c|c|c|c|c|c|c|}
\br Sample &    $L$, &   $w$,  &   $d$, &  $R_{4.2}$, & $R^\square$, &
$T_{\rm c}$, & $l_{i}$, &  $R_{300}$, \\
       &  $\mu$m &  $\mu$m &   nm   &  $\Omega$   &  $\Omega$    &
K       &  nm      &  $\Omega$   \\
\mr SnW5   &  92     &   42    &    120 &  0.140      &  0.064 &
3.789  &   145   &   2.270 \\

SnW12  &   90     &   18    &    332 &  0.038      &  0.008       &
3.836  &   466   &   1.880 \\
\br
\end{tabular}
\end{center}
\end{table}

\begin{figure}
\includegraphics[width=0.5\textwidth]{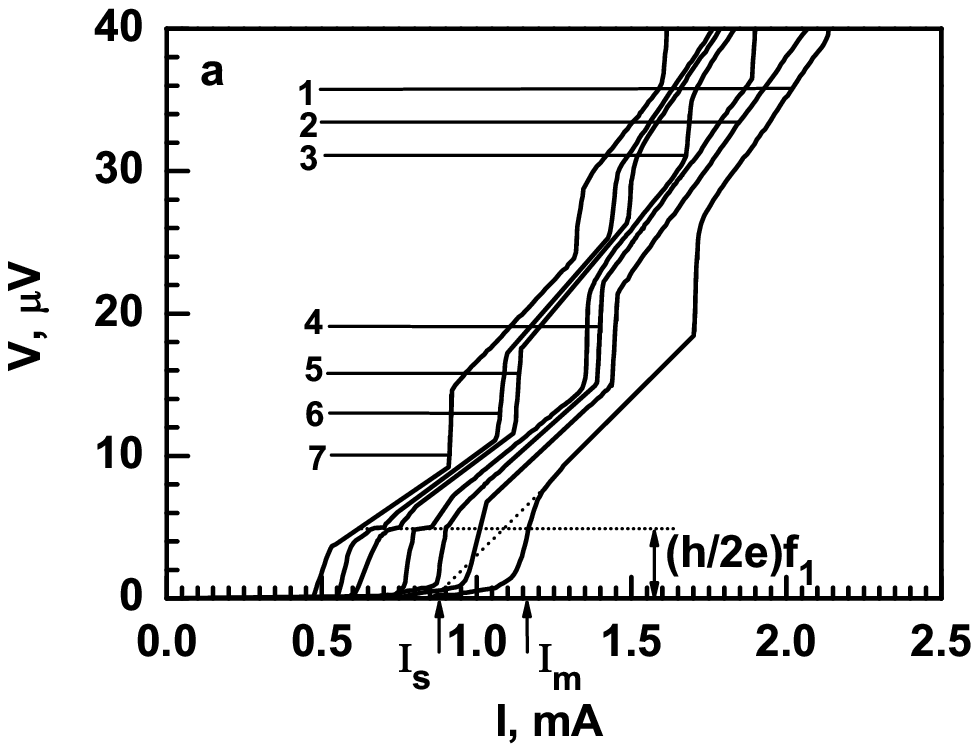}
\includegraphics[width=0.5\textwidth]{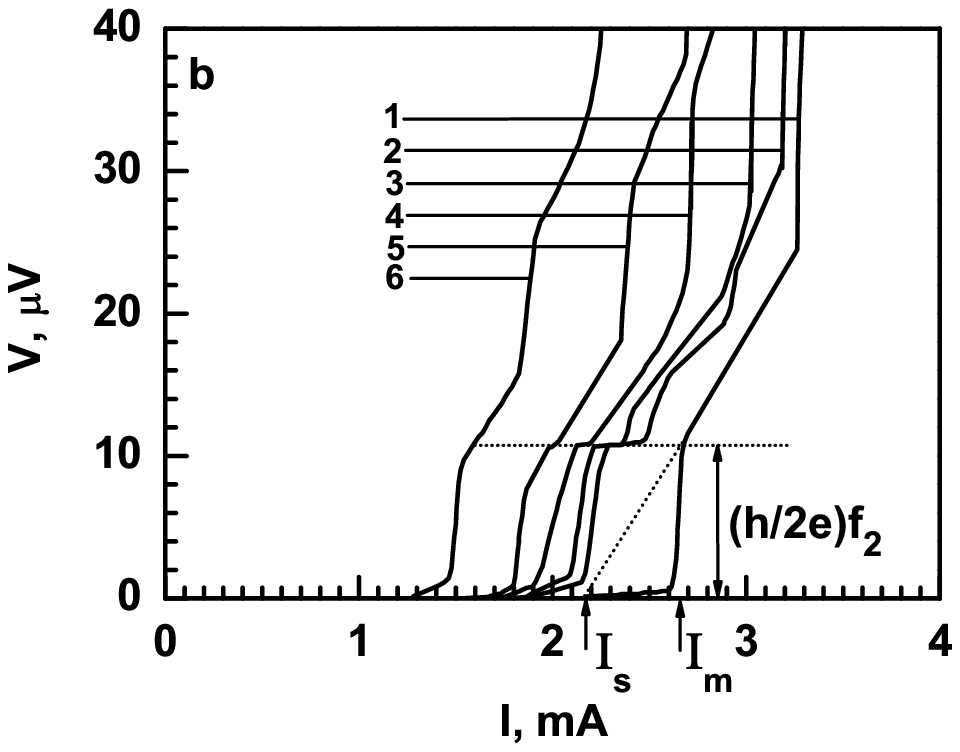}
\caption{
A family of IVCs for the films SnW12 at $T=3.807$ K and $f_1=2476$
MHz (a) and SnW5 at $T=3.744$ K and $f_2=5500$ MHz (b). For curve
1, the applied irradiation power equals zero; for the others it
increases with the IVC number. \label{f} }
\end{figure}

Families of the IVCs for the SnW12 and SnW5 samples, (a) and (b),
respectively, measured at different irradiation power, are
presented in figure \ref{f}. For the first IVC, the rf power is
zero, while for the others it increases with the IVC serial
number. We note that the inequality $w/\lambda_{\perp}(T) \geq 20$
is fulfilled for both films and their IVCs contain initial
resistive regions caused by the vortex motion, i.e., both films
can be unambiguously referred to as wide ones. The IVCs of the
films on initiation of PSLs have the same shape as the IVC of a
narrow channel on initiation of PSCs: they reveal abrupt voltage
steps, cut-off current $I_{\rm s}$ at zero voltage, excess current
at high voltages, and the sample resistance changes by a multiple:
$R=nR_{\rm d1}$, where $R_{\rm d1}$ is the dynamic resistance of a
IVC linear portion corresponding to the first PSL and $n$ is the
number of PSLs in the film. Under the microwave irradiation,
current steps occur in the IVCs at fixed voltage, proportional to
the frequency $f$.

As soon as we suggest the phase slip processes to be similar in a
narrow channel and in a wide film, it would be natural to use the
experience obtained when investigating the ac Josephson properties
of PSCs, considering the PSL as a source of the Josephson
irradiation. Then the observed current steps can be identified as
the Shapiro steps in the IVC of the Josephson irradiation source
at the applied voltages \cite{Shapiro}
\begin{eqnarray} \label{Shapiro}
V= \frac{nh}{2e}f ~~~~~~~~~~(n=1,2, ...),
\end{eqnarray}
which are known to result from the interaction between the applied
microwave field and the ac Josephson current.

We note that the experimental observation of the current steps,
associated with the Josephson irradiation of the PSL, requires
fulfilment of several essential conditions and exclusion of some
side effects. First, the current steps may also appear in the
vortex state, as the result of mutual synchronization
\cite{Anderson} between the applied irradiation and moving vortex
chains. To avoid this, we choose high enough irradiation
frequency, thus adjusting current step position \eref{Shapiro}
inside the voltage region of the dynamic PSL resistance in the
IVC, where the state of the sample is known to be free of vortices
\cite{Dmitriev1}. At the same time, the frequency should be
smaller than the lower frequency of occurrence of the
superconductivity enhancement \cite{Dmitriev2}, which causes both
the critical current and the magnitude of the voltage step on the
PSL initiation to considerably enhance with increasing irradiation
power. We found that in the presence of the enhancement effect,
the current step cannot be observed, because its expected position
falls into the PSL voltage step. This essentially confines maximum
possible value of the irradiation frequency.

In its turn, the constraint to the highest frequency value imposes
limitation on the magni\-tude $V_1$ of the voltage step on
initiation of the first PSL, which must be smaller than the
voltage $hf/2e$ at which the current step occurs. This requires
samples with large enough specific conductance, i.e., with
comparatively large mean free path $l_i$. At first, we studied
rather thick films with a large $l_i \sim 500$ nm, as, for
example, a SnW12 film, in which the magnitude of the first PSL
voltage step is small enough, $V_1 \sim 6~ \mu$V. For thinner
films with smaller $l_i \sim 150$ nm, as, e.g., for a SnW5 film,
we also managed to observe the current step, though we had to
increase the irradiation frequency because of increase in the PSL
voltage step $V_1 \sim 10~\mu$V. As is evident from figure
\ref{f}, even at a maximum possible frequency, the PSL voltage
step in the IVC with no microwave field applied still remains
larger than the expected voltage value for the occurrence of the
current step, $V_{1}>hf/2e$. However, while the irradiation power
increases, the values of the critical current and the voltage step
decrease, providing a possibility for observation of the current
step in the region of the dynamic PSL resistance.

\begin{figure}
\centerline{\includegraphics[height=3in]{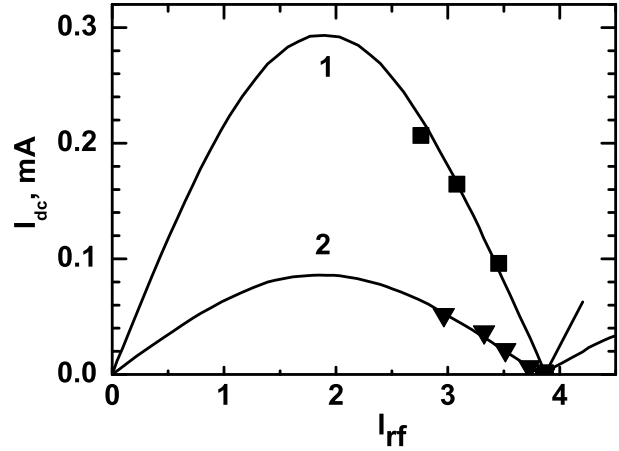}}
\caption{
The amplitude $I_{\rm dc}$ of the first current step in the IVC as
a function of the amplitude of the rf current $I_{\rm rf}\propto
\sqrt{P}$ in units of $hf/2eR_{\rm d1}$:

\noindent SnW5 sample -- $\blacksquare$, $T=3.744$ K, $f=5500$
MHz, $R_{\rm d1}=0.018$ $\Omega$, $I_{\rm J}=0.259$ mA;

\noindent SnW12 sample -- $\blacktriangledown$, $T=3.807$ K,
$f=2476$ MHz, $R_{\rm d1}=0.015$ $\Omega$, $I_{\rm J}=0.075$ mA.

\noindent Curves 1 and 2 - theoretical results.
 } \label{f2}
\end{figure}

Figure \ref{f2} shows the dependence of the first current step
$I_{\rm dc}$ on the rf current $I_{\rm rf}\propto \sqrt{P}$. For
the dc voltage source, the dependence of the current step
amplitude on the applied ac voltage is known to follow the Bessel
function \cite{Grimes}. However, in practice, both the impedance
of a microwave oscillator and the resistance of the dc source are
usually larger than the impedance of the Josephson oscillator. In
our experiments, the dynamic resistances of the PSL were $R_{\rm
d1}=0.018$ Ohm for SnW5 and $R_{\rm d1}=0.015$ Ohm for SnW12,
i.e., much smaller than the resistances of the microwave
oscillator ($50$ Ohm) and the dc current source ($470$ Ohm). Since
the sizes of our samples are small compared to the electromagnetic
field wavelength (the film length is $\sim 10^{-4}$ m and the
maximum wavelength is $\sim10^{-2}$ m), we suggest that the
microwave irradiation generates a spatially homogeneous rf current
passing through the sample, $I_{\rm rf}\propto \sqrt{P}$ ($P$ is
the irradiation power), although its absolute value was not
measured.

In such a case, the assumption of given current applied to a PSL
will be more adequate, and the equation of the resistive model of
the Josephson junction \cite{Cumber1,Stewart} can be used for
description of the time dependence of the phase difference
$\varphi$. For low-capacitance junctions, this equation reads
\begin{eqnarray} \label{Barone}
\frac{I_{\rm rf}}{I_{\rm J}}\sin{\Omega \tau}+\frac{I_{\rm
dc}}{I_{\rm J}}=\frac{d\varphi}{d\tau}+\sin{\varphi}.
\end{eqnarray}
Here the dimensionless time and frequency are given by
$\tau=\frac{2eR_{\rm d1}I_{\rm J}}{\hbar}t$ and
$\Omega=\frac{hf}{2eR_{\rm d1}I_{\rm J}}$; $I_{\rm J}$ is the
maximum Josephson current passing through the junction.

Theoretical dependencies of the current step $I_{\rm dc}$ on the
amplitude of $I_{\rm rf}$ obtained by numerical solution of
\eref{Barone} are shown in figure \ref{f2} by solid lines. As
noted in \cite{Russer}, the time evolution of the phase $\varphi$
under the microwave irradiation, and hence the behaviour of the
current step, essentially depends on the dimensionless frequency
$\Omega$: at $\Omega \gg 1$, the dependence of the step amplitude
on the rf power, obtained in a framework of the resistive model
\eref{Barone}, approaches the result of the given voltage model
\cite{Grimes}. In our case, the dimensionless frequency is rather
high, $\Omega=2.44$ for SnW5 and $\Omega=4.55$ for SnW12, which
indicates the validity of the both models for the description of
our experimental data. Numerical simulations for these particular
values of $\Omega$ also confirm this conclusion.

In figure \ref{f2}, the rf current amplitude is expressed in units
of $hf/2eR_{d1}$. Since only relative changes in $I_{\rm rf}$ can
be measured, the absolute values of $I_{\rm rf}$ were evaluated
assuming the theoretical value of $I_{\rm rf}$ to be equal to the
experimental one at the point where $I_{\rm dc}=0$. Another
adjustable parameter is the amplitude of the Josephson current
$I_{\rm J}$ through the PSL: by fitting the theoretical curve
$I_{\rm dc}(I_{\rm rf})$ to the experimental points, we obtain
$I_{\rm J}=0.26$ mA for the SnW5 sample and $I_{\rm J}= 0.075$ mA
for SnW12. Such low values of the superconducting current compared
to the dissipative and critical currents have been also suggested
in the theoretical PSL model \cite{Lempitskij}. Estimations of the
Josephson irradiation power from the PSL by formulae $P=(I_{\rm
dc}^{\rm max})^2 R_{\rm d1}$ \cite{Kuznechsov} ($I_{\rm dc}^{\rm
max}$ is the maximum magnitude of the current step) and $P=I_{\rm J}^2
R_{\rm d1}$ give almost the same results: $P_{\rm
SnW12}\approx10^{-10}$ W for SnW12 and $P_{\rm
SnW5}\approx10^{-9}$ W for SnW5, in agreement with the Josephson
irradiation power from PSC \cite{Kuznechsov} within an order of
magnitude.

Since the current step occurs at a certain finite value of the rf
power when the lower edge of the linear IVC portion approaches the
voltage value $hf/2e$ (see figure \ref{f}), only a descending
branch of the whole dependence $I_{\rm dc} (I_{\rm rf})$ was
detected in the experiment, as obvious from figure \ref{f2}. We
also note that a part (from initiation to disappearance) of only
one ``period'' of the expected oscillating dependence
\cite{Grimes} of the current step on the irradiation power was
observed; the step does not reappear while the rf power further
increases. Similar effect was found in the study of current steps
in the IVCs of whiskers with PSCs \cite{Kuznechsov}; the reason of
this is unknown yet.

In our experiments, we managed to detect only the first (main)
Shapiro step correspon\-ding to the integer $n=1$ in
\eref{Shapiro}. The absence of the steps associated with higher
harmonics, $n>1$, is a common feature of the experiments on the
superconducting films \cite{Skocpol} and whiskers
\cite{Tidecks,Kuznechsov}. The authors of these papers have no
explanation for this fact. In our case we suppose, that the most
obvious reason for this is that the expected positions of the
higher current steps fall into the PSL voltage steps, as seen from
figure \ref{f}(b) for the sample SnW5; for a similar reason, the
steps with $n=2,3$ were not observed in \cite{Tidecks}. For the
sample SnW12 (see figure \ref{f}(a)), the current step at $n=2$
might be in principle detected at the linear IVC portion; however,
as follows from the analysis in \cite{Grimes}, the amplitude of
the $n$-th step decreases with increase in $n$ at fixed magnitude
of the rf power. Thus, due to relatively small value of the main
step in this case, the value of the second step may appear beyond
the accuracy of our measurements. It should be also noted that the
intrinsic dynamics of the weak link in a PSC or a PSL is much more
complicated and considerably affected by the microwave power, in
contrast to that of artificial (`hand-made') Josephson junctions.
Thus a more reasoned explanation of the experimental deviations
from the results of the traditional theory of the Shapiro steps
requires creation of a consistent theory of the self-organized
Josephson weak links like PSCs and PSLs under external
irradiation.

\section{Conclusions}

In conclusion, current steps resulted from the interaction between
the intrinsic Josephson irradiation of phase slip lines and the
applied electromagnetic field were first observed in the
current-voltage characteristics of deliberately wide
superconducting films, whose transition to the resistive state
with phase slip lines is preceded by creation of a vortex state.
The dependence of the current step magnitude on the rf power is
similar to the behaviour of the current steps, associated with the
phase slip centers in narrow superconducting channels. This gives
the experimental evidence that the nonstationary properties of
phase slip lines and phase slip centers are largely identical.

\section{Acknowledgments}

The authors express their thanks to Salenkova T.V. for technical assistance
and Khristenko E.V. for helpful discussions.

\end{document}